\documentclass{aa}  
\RequirePackage{nameref}
\usepackage{graphicx}
\usepackage{txfonts}
\usepackage{lineno}
\usepackage{setspace}
\usepackage{soul}

\begin{document}


\title{Three dimensional magnetic reconnection mediated with plasmoids and the resulted multi-thermal emissions in the cool atmosphere of the Sun}
\subtitle{}
   \author{Guanchong Cheng
          \inst{1,2,3}
          \and
          Lei Ni\inst{1,3,4}
                    \and
          Robert Cameron\inst{2}
                    \and
          Hardi Peter\inst{2}
          \and
          Yajie Chen \inst{2}
          \and
          Jun Lin\inst{1,3,4}
          }

   \institute{Yunnan Observatories, Chinese Academy of Sciences,Kunming, Yunnan 650216, P. R. China\\
              \email{leini@ynao.ac.cn,cameron@mps.mpg.de}
         \and
	    Max Planck Institute for Solar System Research, Justus-von-Liebig-Weg 3, 37077, Göttingen, Germany
         \and
             University of Chinese Academy of Sciences, Beijing 100049, P. R. China
         \and
			Yunnan Key Laboratory of Solar Physics and Space Science, Kunming 650216, PR China}


 
  \abstract
  {Flux emergence is ubiquitous in the Sun's lower atmosphere. The emerging flux can reconnect with the pre-existing magnetic field.}
  {We aim to investigate plasmoid formation and the resulted multi-thermal emissions during the three dimensional reconnection process in the lower solar atmosphere. }
  {We conducted 3D radiation magnetohydrodynamic (RMHD) simulations using the MURaM code, which incorporates solar convection, radiation. We simulated the emergence of a flat magnetic flux sheet which was introduced into the convection zone.  For comaprison with results previously reported from observations, we employed the RH1.5D code to synthesize H$\alpha$ and  Si IV spectral line profiles, and the ultraviolet images are synthesized by using the optical thin methods.}
  {Flux emergence took place as part of the imposed flux tube crossed the photosphere. In the lower solar atmosphere, magnetic reconnection occurred, forming thin, elongated current sheets. Plasmoid-like features appear as part of the reconnection process, resulting in many small twisted magnetic flux ropes, which are expelled toward the two ends of the reconnection region.
    Consequently hot plasma with a temperature exceeding 20,000 K and much cooler plasmas with a temperature below 10,000 K can coexist in the reconnection region.
    Synthesized images and spectral line profiles through the reconnection region display typical characteristics of reconnection occuring in the lower solar atmosphere, such as Ellerman bombs (EBs) and UV bursts. The cooler
plasmas showing characteristics of EBs can
be found above hot plasma, reaching altitudes more than 2Mm above the solar surface. Meanwhile, some hot plasma featuring characteristics of UV bursts can extend downwards to the lower chromosphere, approximately 0.7Mm above the solar surface.  }
  {Our simulation results indicate that the turbulent reconnection mediated with plasmoid instability can occur in small-scale reconnection events
    such as EBs and UV bursts. The coexisting of hot and much cooler plasmas in such a turbulent reconnection process can well explain the UV bursts temporally and spatially connecting with EBs. 
  }
 \maketitle{}
  \titlerunning{}
  \authorrunning{}
%
\section{Introduction}

In the lower solar atmosphere, numerous small-scale brightenings at different wavelengths have been observed
\citep[e.g.,][]{Katsukawa2007,Peter,Huang2017,Toriumi2017,Young,Rouppe,Joshi2022}.
Magnetic reconnection is believed to be one of the main mechanisms driving these events.
In the past few decades, many works studied these events, including the characteristics of spectral line profiles
and images at different wavelength pass bands \citep[e.g.,][]{Fang2006,Pariat2007,Peter,Tian2016,Leenaarts2025}.
The corresponding magnetic field structures \citep[e.g.,][]{Pariat2004,Toriumi2017,Chitta2017a,Chitta2017b,Samanta2019}  are potentially important to understand their formation process and heating mechanisms.

Limited by the resolution of current observational equipments, further investigation into the fine structures and formation processes of EBs and UV bursts are significantly challenging.  In this context, numerical simulations play a crucial role. Early simulations confirmed that the reconnection processes triggered by U-shaped magnetic field lines, due to Parker instability,
are key processes in the formation of EBs \cite[e.g.,][]{Isobe2007,Archontis2009}.
More recent radiative MHD (rMHD) simulations investigated the formation process of EBs, including radiative transfer process,
from the upper convection zone to the lower solar atmosphere.
For instance, \cite{Danilovic2017} utilized the MURaM code \citep{Vogler2005} to study the reconnection process in the photosphere. 
Those works carefully analyzed the whole evolution process and synthesized images and spectral line profiles from different line of
sights, and showed the typical characteristics of EBs. \\


Numerical simulations of magnetic reconnection in the partially ionized low solar atmosphere show that the plasmas can be heated to a high temperature above 20,000 K when the reconnection magnetic field is strong enough and the plasma $\beta$ is smaller than 0.1 \cite[e.g.,][]{Ni2015,Ni2016,Peter2019}.  The RMHD simulations by \cite{Hansteen2017} studied small scale magnetic reconnection in the flux emergence region, their results indicated that the UV bursts are generated in a current sheet extended from the middle chromosphere to the transition region. But the later simulations by \cite{Ni2022} showed that the strong Si IV emissions can also be generated in the lower chromosphere as long as the reconnection magnetic fields are stronger than 500 G, the radiative cooling model used in \cite{Ni2022} is similar as that in \cite{Hansteen2017} for the chromosphere.  \\

Recently,  the UV bursts connecting with EBs have been studied in several numerical works. The 3D RMHD simulations by \cite{Hansteen2019} found a vertical current sheet with a hot upper part and a much cooler lower part, which are corresponding to UV busts and EBs, respectively.  They proposed that UV bursts and EBs can be generated in the same reconnection process, but these two brightenings are located at different atmospheric layers and they are far apart from each other \citep{Ada2020}. In contrast, \cite{Ni2021} performed high resolution 2.5 D simulations of magnetic reconnection between emerged and background magnetic fields. Their results showed that the plasmoid instability causes the highly nonuniform density distributions in the curved current sheet, the low-density high-temperature (>20,000 K) and high-density low-temperature (<10,000 K) plasmas alternatively appear in the space, the potential hot UV emissions and much cooler H$\alpha$ wing emissions can locate approximately at the similar height, even within the same plasmoid in the lower chromosphere. Their recent work further confirmed this model by using the radiative transfer code RH1.5D to synthesize different spectral line profiles and diagnose the different emission sources, the more realistic radiative cooling and ionization effects are included in \citep{Cheng2024}. \\

%

\begin{figure*}[ht]
	\centering
	\includegraphics[width=16cm]{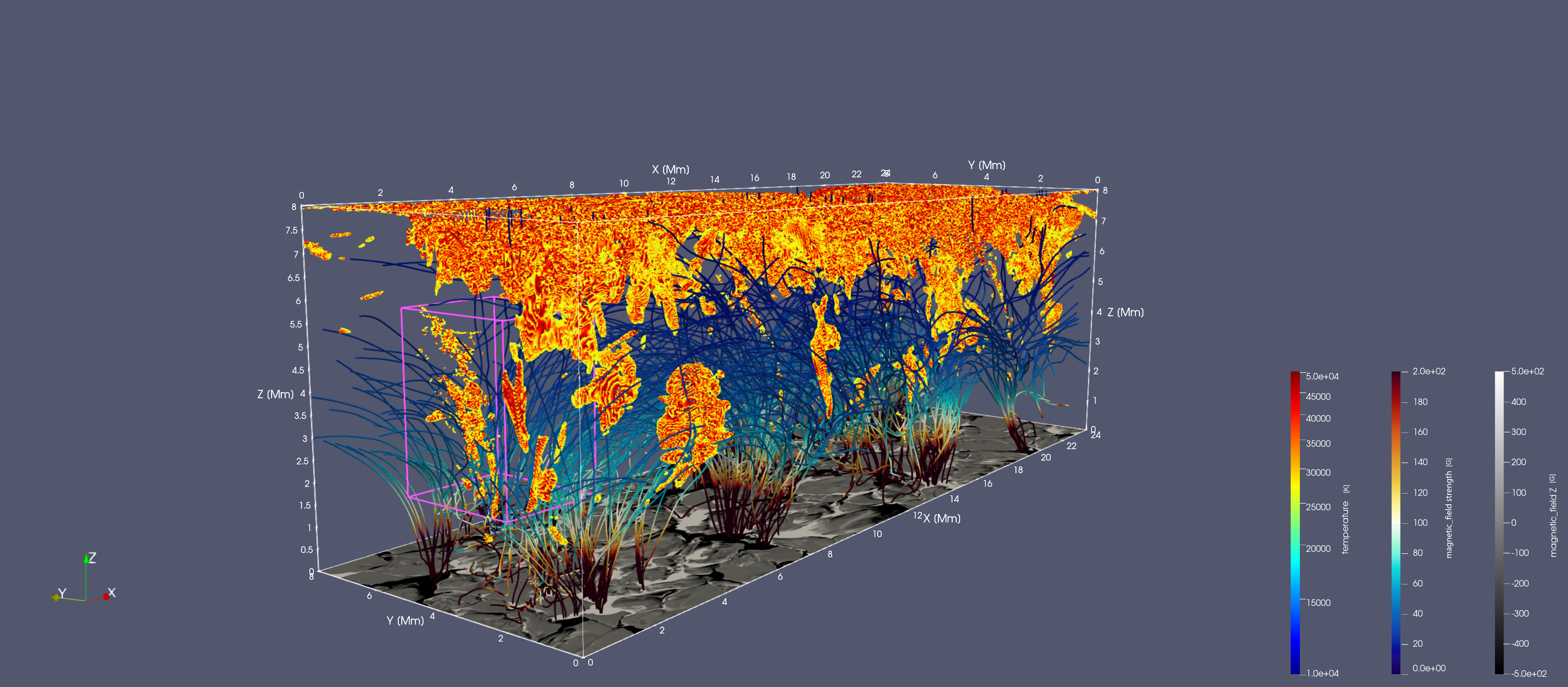}
	\caption{3D overview of the simulation domain above the photosphere at t=4414s: The colorful line structures represent the magnetic field lines in the simulation area, with colors corresponding to the field strengths. The grayscale slice at the bottom illustrates the distribution of the longitudinal magnetic field at the solar surface. Yellow and red isosurfaces represent the regions where temperatures are above 20,000 K. The purple box represents the region of interest (ROI). More details can be found in the supplementary movies.}
	\label{fig:Setup}%
\end{figure*}

\begin{figure*}[ht]
	\centering
	\includegraphics[width=18cm]{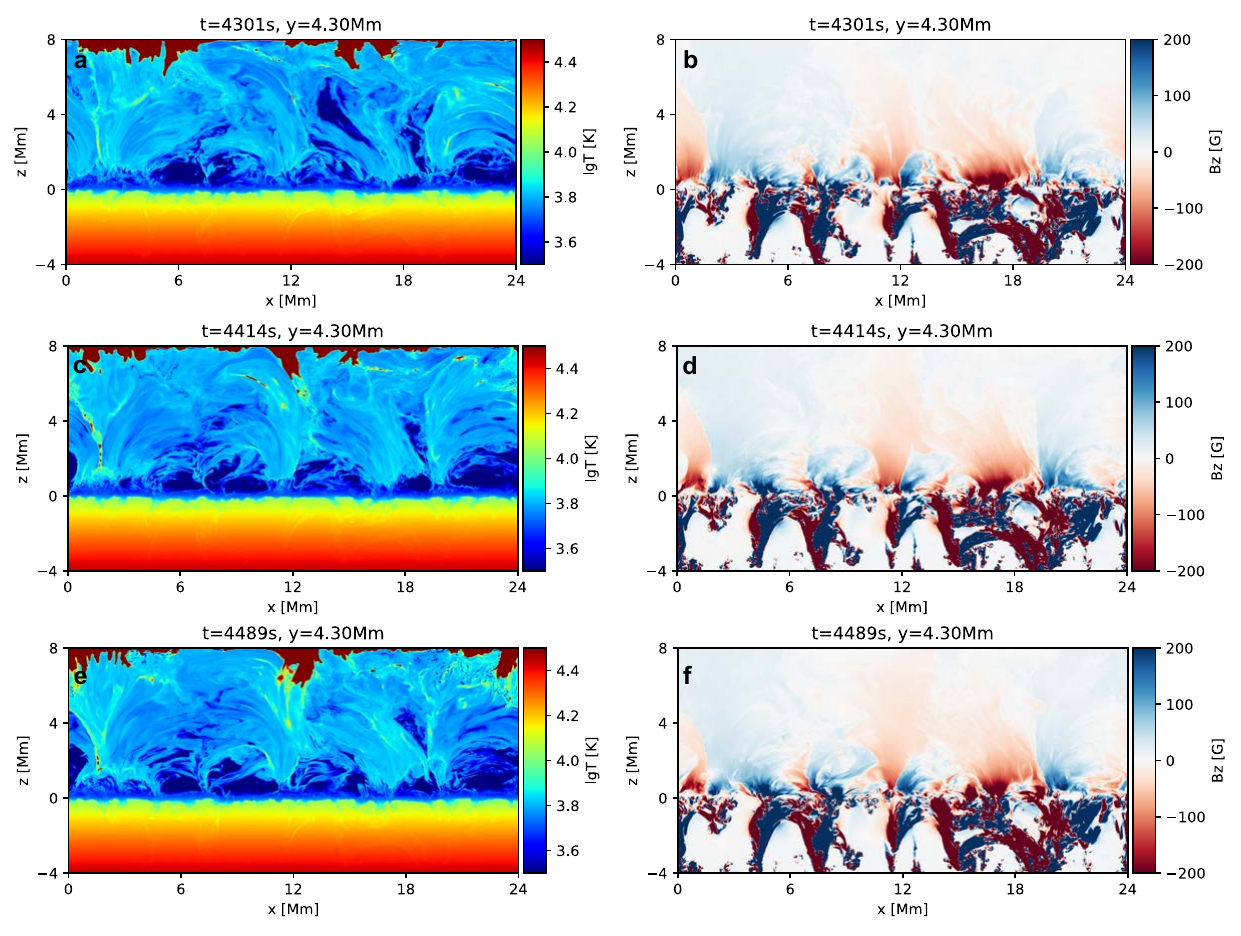}
	\caption{Slices at $y = 4.3$ Mm at three different times in the simulation. The left column displays the temperature distribution in the slice planes, while the right column shows the distribution of the magnetic field strength (Bz) perpendicular to the solar surface. More details can be found in the supplementary movies.}
	\label{fig:Slices1}%
\end{figure*}

\begin{figure*}[ht]
	\centering
	\includegraphics[width=18cm]{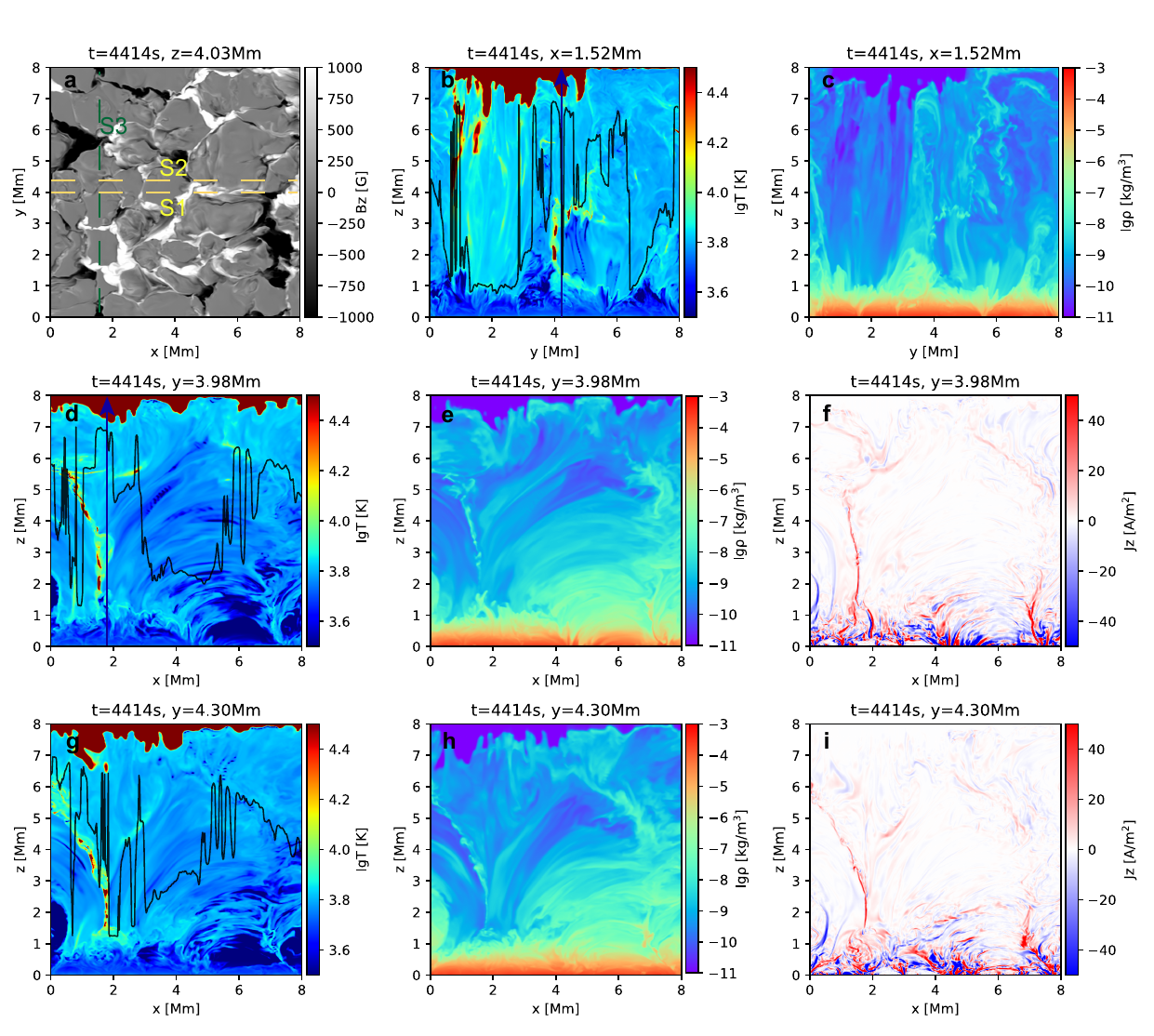}
	\caption{Slices along different directions at $t = 4414$ s in the simulation: Panel a illustrates the longitudinal magnetic field (Bz) at the solar surface, marked with dashed lines to indicate the positions of slices S1, S2, and S3. Panels b and c represent the temperature and density distributions along slice S3 within the y–z plane. Panels d to f and panels g to i respectively show the temperature, density, and current density (Jz) distributions along slices S1 and S2 in the x–z plane. The black curves in the figure indicate the height at which the H$\alpha$ line center optical depth equals unity.}
	\label{fig:Slices2}%
\end{figure*}

\begin{figure*}[ht]
	\centering
	\includegraphics[width=14cm]{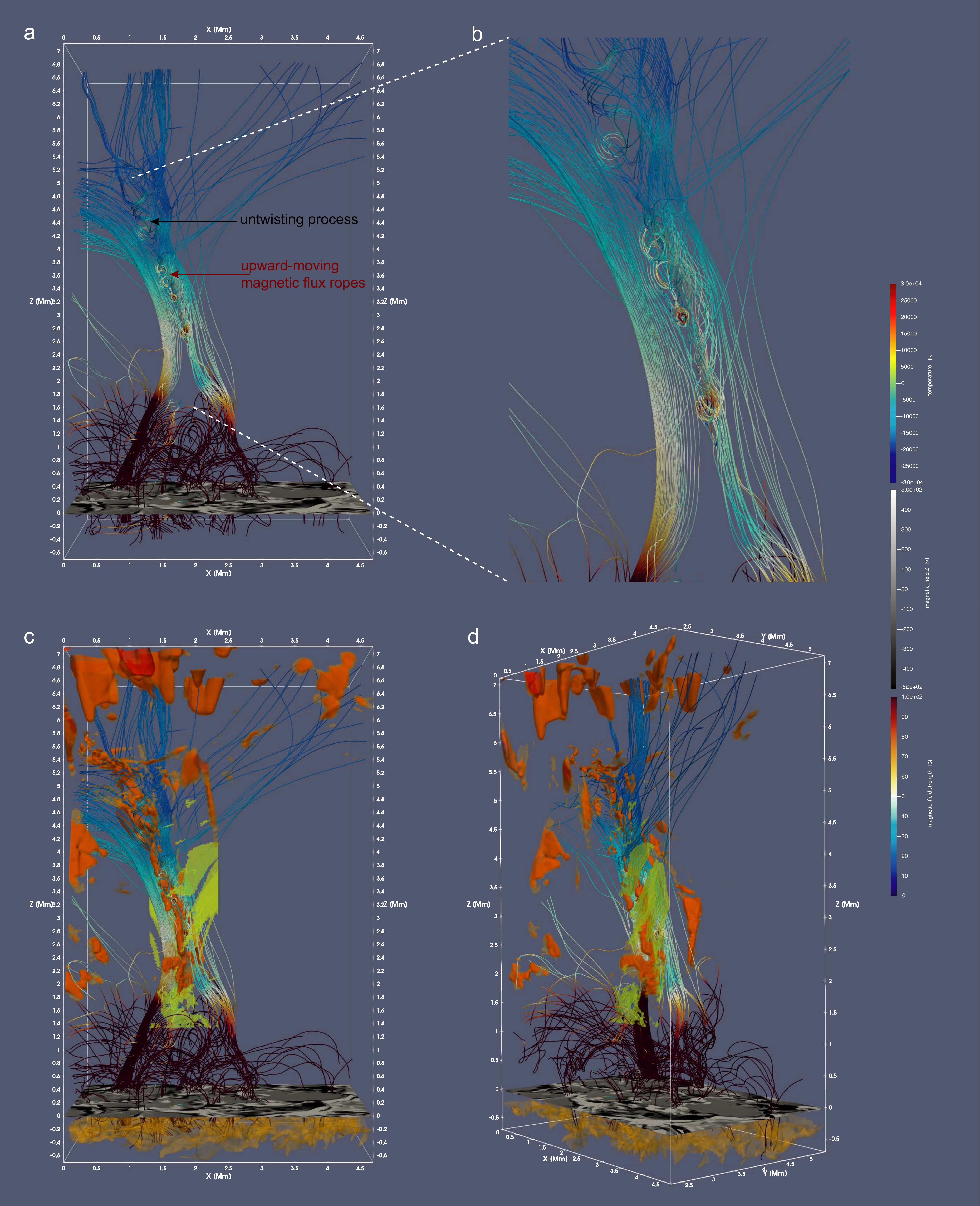}
	\caption{Three-dimensional views of the interesting region at t=4414 s: The grayscale image at the bottom represents the longitudinal magnetic field at the solar surface. Lines with different colors indicate magnetic field lines. The reddish isosurfaces indicate the regions with high-temperature plasma, while greenish isosurfaces represent low temperature regions. Panel b is a zoomed-in view of the reconnected region in Panel a. More details can be found in the supplementary movies.}
	\label{fig:plasmoid}%
\end{figure*}

\begin{figure*}
	\centering
	\includegraphics[width=16cm]{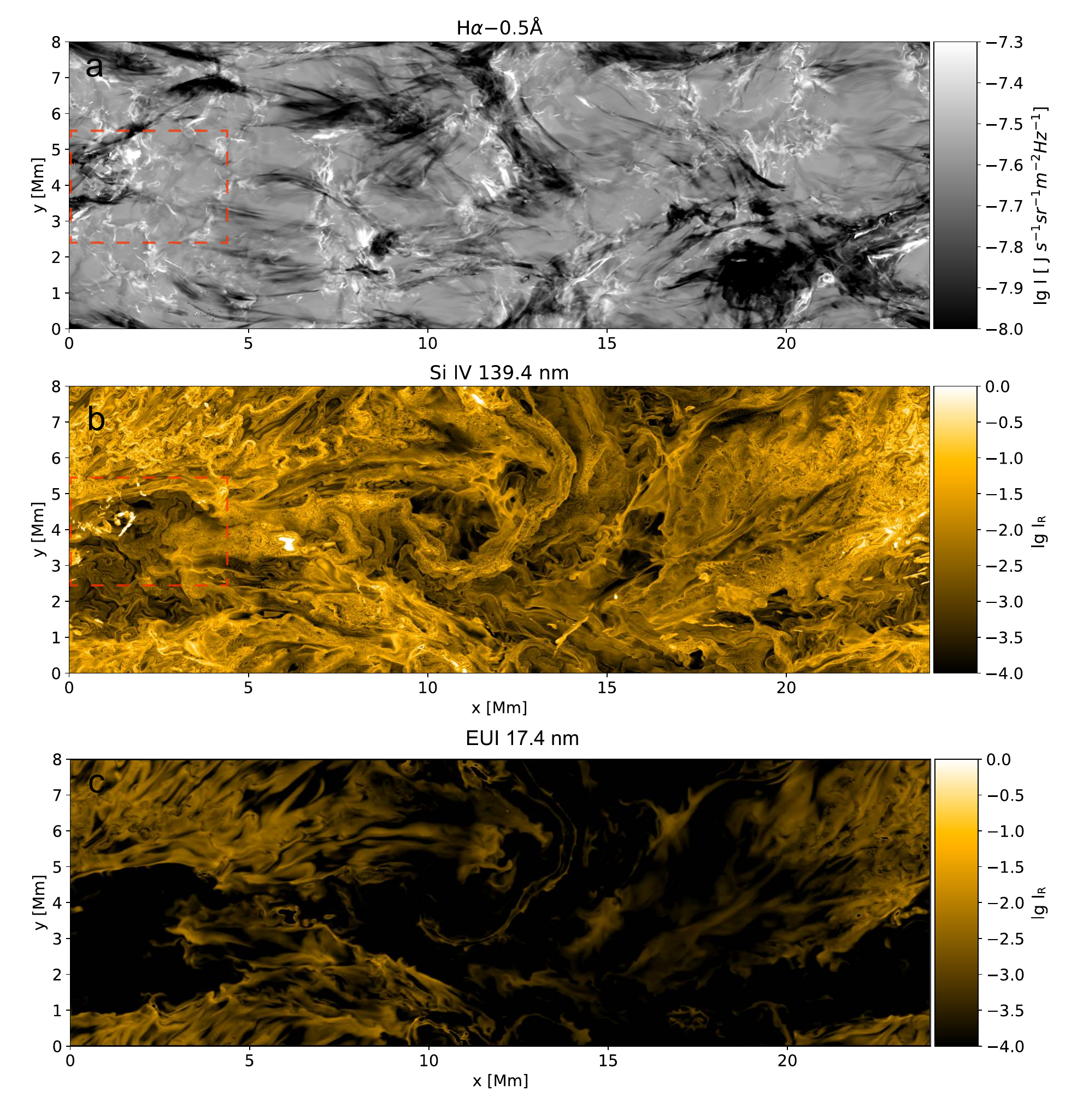}
	\caption{Synthetic observables from a line-of-sight oriented vertically to the solar surface at t=4414s in the simulation: Panel a displays the H$\alpha$ wing radiation at the H$\alpha$-0.5Å wavelength calculated by using the RH1.5D code. Panel b shows the Si IV 139.4 nm synthetic radiation image calculated by using the optically thin method. Panel c presents the synthetic image of the EUI 17.4 nm band, also calculated by using the optically thin method. Hereafter, $\mathrm{I}_{\mathrm{R}}$ denotes the relative radiation intensity.}
	\label{fig:syn1}%
\end{figure*}

\begin{figure}[ht!]

	\includegraphics[width=9cm]{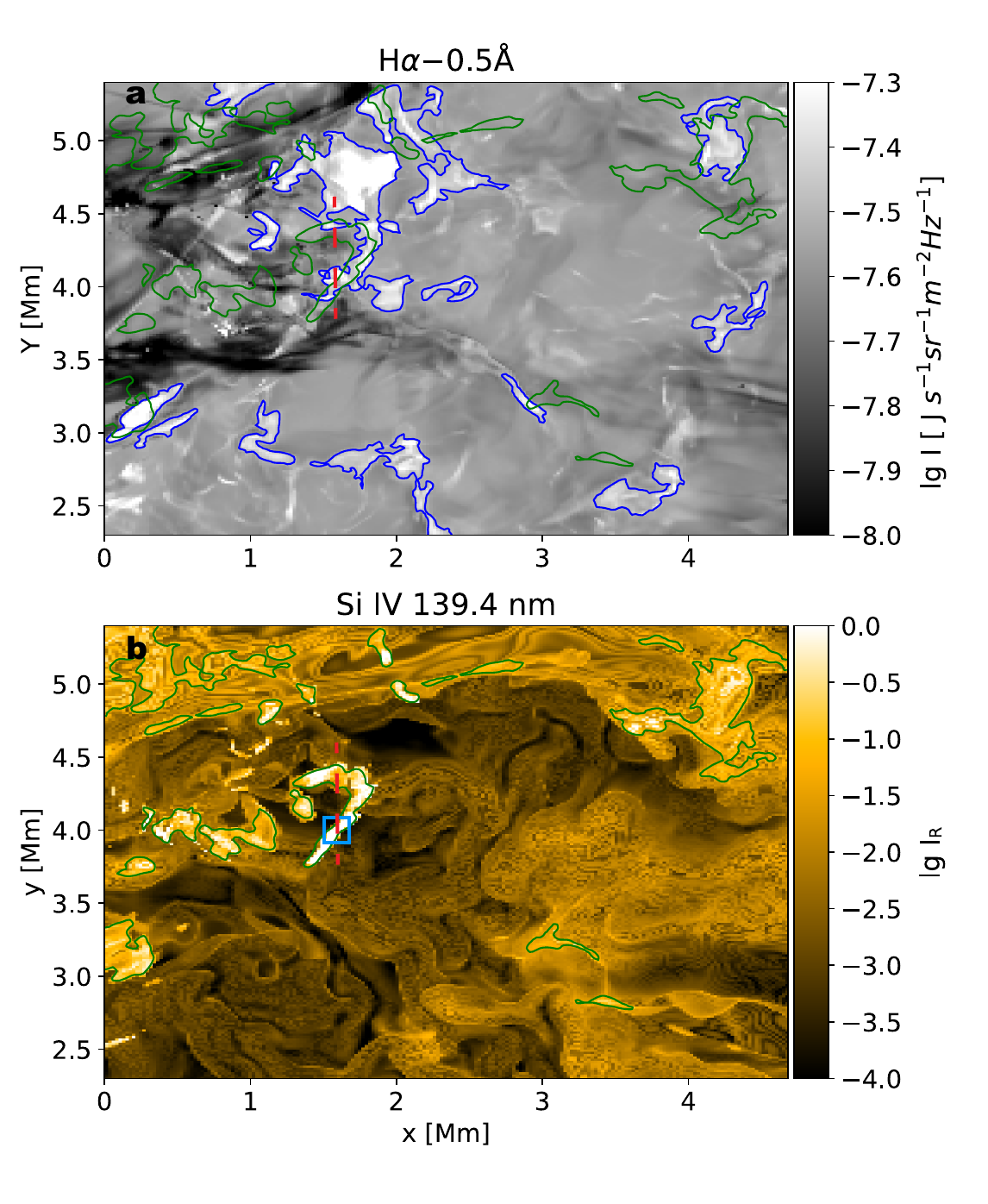}
	\caption{
		Synthetic observables in the red dashed rectangular area of Fig.~\ref{fig:syn1} .	}
	\label{fig:synbox}%
\end{figure}

\begin{figure*}

	\includegraphics[width=18cm]{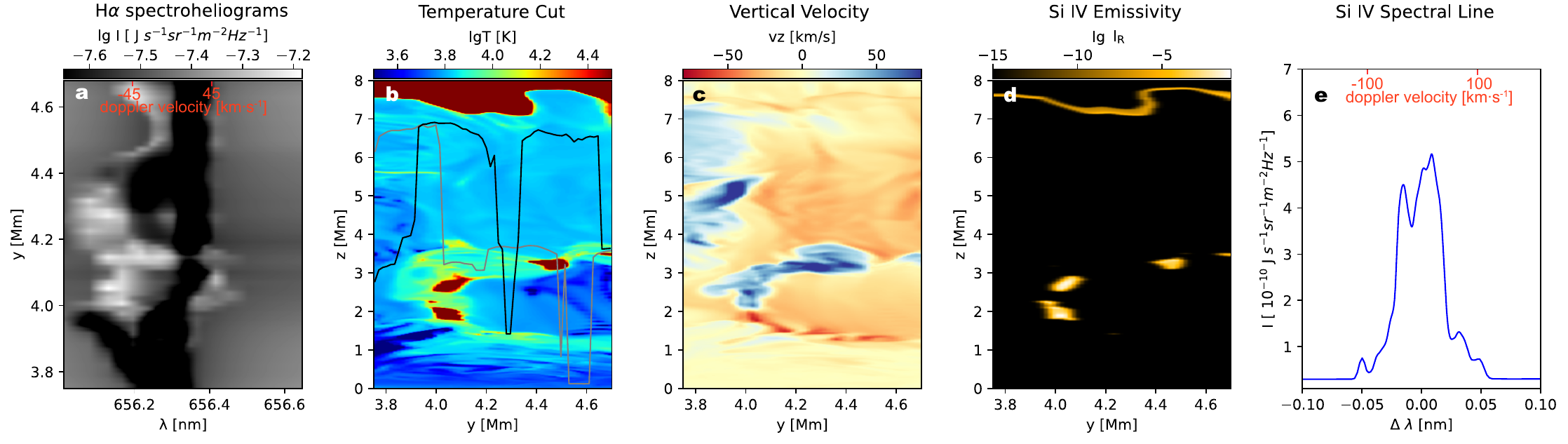}
	\caption{Panel a shows the synthetic H$\alpha$ spectrum at t = 4414 s. It displays the synthesized spectrum for comparison with Ellerman's observations from 1917, with the horizontal axis representing wavelength and the vertical axis representing the length along the red dashed line in Fig.~\ref{fig:synbox}. Note: The red dashed line in Fig.~\ref{fig:synbox}a acts as an imaginary spectrograph slit. Panels b to d present physical quantities in the y–z plane along the red dashed line indicated in Fig.~\ref{fig:synbox}. In panel b, the temperature distribution is shown, with the gray and black lines representing the locations where the optical depths of the H$\alpha$ wing (-0.5 Å) and H$\alpha$ line center reach unity, respectively. Panel c displays the vertical velocity (vz) distribution, and panel d shows the emissivity of Si IV, and panel e displays the synthesized Si IV spectral line profile calculated from the blue rectangular region marked in Fig.~\ref{fig:synbox}b. }
	\label{fig:spectrohelio}%
\end{figure*}

\begin{figure*}
	\includegraphics[width=18cm]{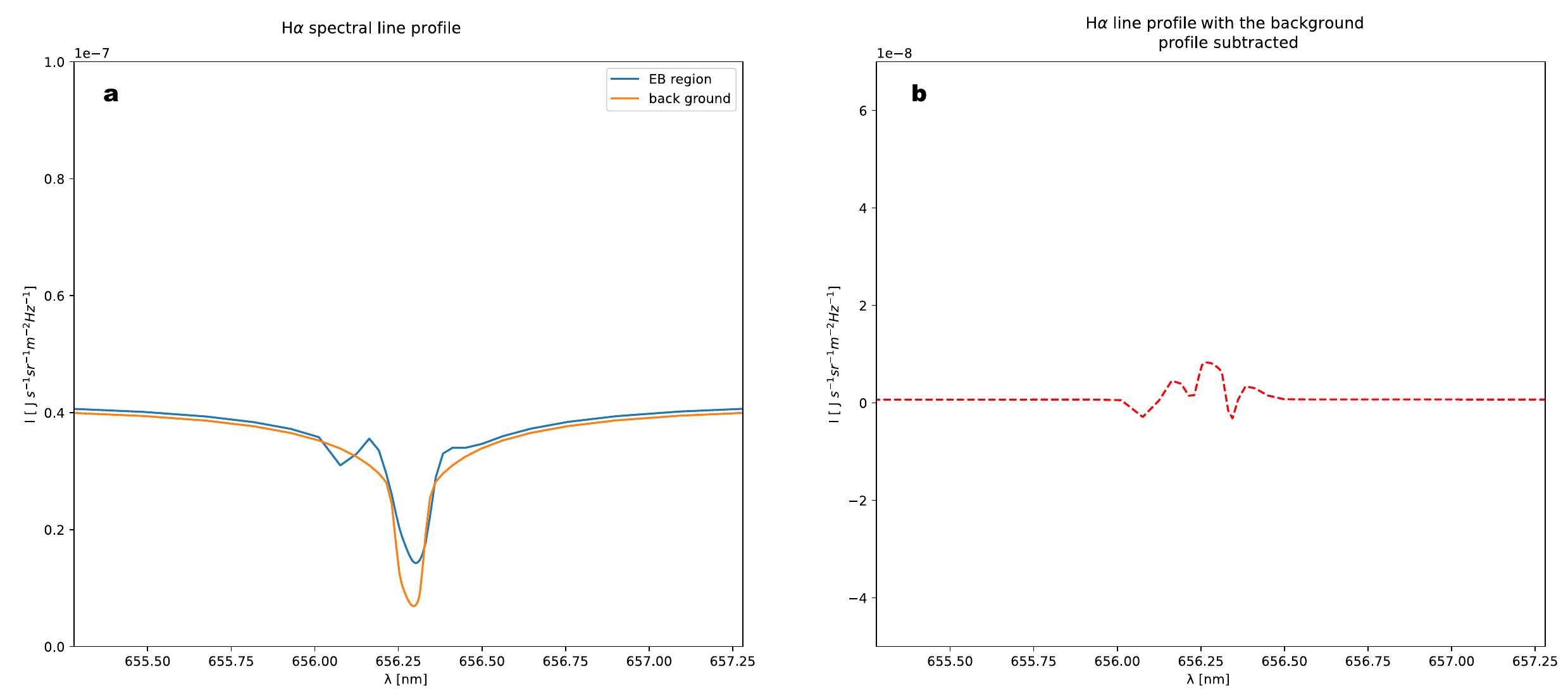}
	\caption{H$\alpha$ spectral line profiles through the low-temperature plasma blob indicated by the blue arrows in Fig.~\ref{fig:Slices2}a\&d. Panel a shows the spectral line profile calculated using RH1.5D directly, and Panel b shows the result after subtracting the background spectral line profile.}
	\label{fig:Halpaline}%
\end{figure*}

\begin{figure*}[ht]

	\includegraphics[width=18cm]{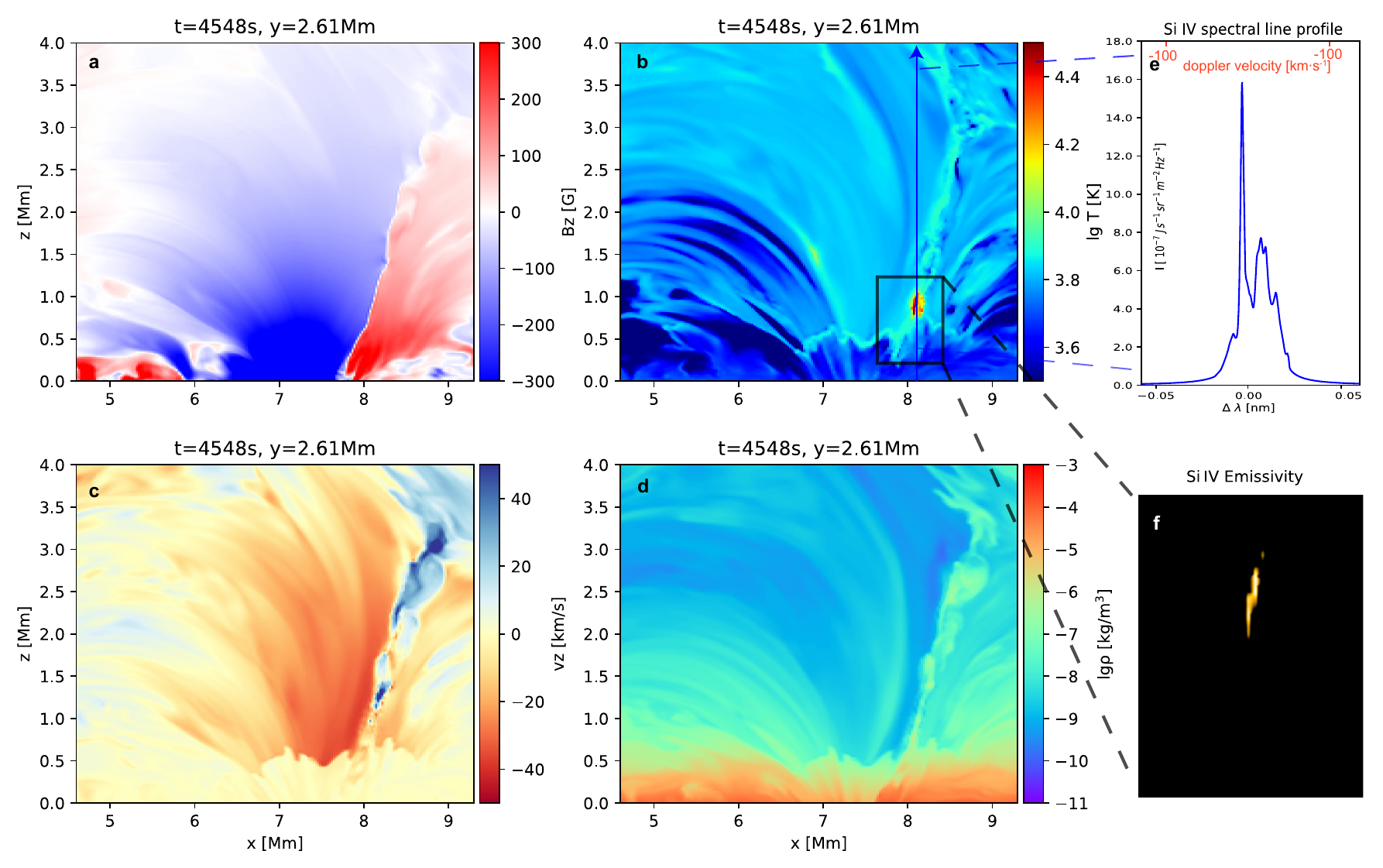}
	\caption{Slices at $t = 4548$ s and $y = 2.61$ Mm of another reconnection event, displaying the distribution of magnetic field in the z-direction, temperature, velocity in the z-direction (Vz), density, the synthesized Si IV spectral line profile, and the synthetic image of the high-temperature area in the Si IV band along the y-direction.}
	\label{fig:slices3}%
\end{figure*}

There is limited observational evidence of plasmoid instability in the lower solar atmosphere, e.g. a few works showing blob-like structures in Ca II K and H$\alpha$ wing images \citep{Rouppe van der Voort2017,Rouppe van der Voort2024}. Recently, \cite{Cheng2024b} have identified the plasmoid-like structures at H$\alpha$ wing images in a small current sheet, and they were ejected downward to cause the brightenings at the post flare loop region. The turbulent reconnection mediated with plasmoids in the partially ionized low solar atmosphere has been presented in many previous 2D MHD simulations \citep[e.g.,][]{Leake2012,Murphy2015,Ni2015,Rouppe van der Voort2017,Ni2018,Peter2019,Ni2021,Liu2023,Wargnier2023}. However, the plasmoids like structures have never been shown in previous 3D RMHD simulations of EBs and UV bursts \citep[e.g.,][]{Hansteen2017,Danilovic2017,Hansteen2019}, except for one recent work by \cite{Cheng2021} which did not include the realistic radiative cooling effect. 

In this work, the advanced RMHD code MURaM was applied to investigate magnetic reconnection triggered by flux emergence in partially ionized lower solar atmosphere. The radiative transfer process results in the realistic plasma and magnetic environment. For the first time, we show that the plasmoid instability appears in most of the small scale reconnection processes relating with EBs and UV bursts in the 3D RMHD simulation. We synthesize the spectral line profiles and images in different wave lengths, and the results further confirm that the hot Si IV emissions and much cooler H$\alpha$ wing emissions can alternately appear in space in a turbulent reconnection region. The remaining part of this paper is organized as follows: Section 2 introduces the numerical model and initial setups. Section 3 presents numerical results. The summaries and discusses are given in Section 4.\\

\section{Numerical setup}

In this study, we utilized the extended MURaM code for our numerical simulations, incorporating capabilities for modeling the coronal and chromospheric extensions. We use the version described in \cite{Rempel2017}, which includes a coronal radiation cooling model and resolves the limitations on time steps caused by high Alfvén speeds, with dissipation settings identical to those described in \cite{Przybylski2022}.
The size of the simulation box is 24 Mm $\times$ 8 Mm $\times$ 12 Mm in the x, y, and z directions respectively, with the z-axis perpendicular to the solar surface.
The upper convection zone extends from z=-4 Mm to z=0 Mm, and the solar atmospheric region extends from z=0 Mm to z=8 Mm. The grid size in x-direction is 23.4 km, and it is 15.63 km in y and z direction. The grid number is 1024 $\times$ 512 $\times$ 768. This resolution is adequate for resolving the small structures within current sheets, even though it is still much lower than that in our previous 2D simulations by employing adaptive mesh refinement (AMR) technology \citep[e.g.,][]{Ni2021,Cheng2024} .\\

In our simulations, we use the LTE radiative cooling model for the photosphere and lower chromosphere, and the optically thin radiative cooling model is applied for the upper chromosphere and corona region.
We begin with a convecting simulation which extends to z=2 Mm. We insert a magnetic flux sheet at z=-2 Mm and the peak strength is 2200 G.  The strength of the magnetic filed decrease with height, and the effective radius is about 0.5 Mm. The convection motions cause the magnetic fields to emerge upward into the solar atmosphere.
When the system reaches a relaxed state, we extend the simulation into the region above z=2 Mm.  The magnetic fields from z= 2Mm to z=8 Mm at the beginning of the second stage is givn by the potential extrapolation method, while the initial temperature is expanded into the corona region by using a tanh function to match to a coronal temperature of {\bf{$10^{6}$}}~K. The initial density and internal energy distributions are derived from the hydrostatic assumption. The velocity above z=2 Mm is set to zero at the beginning of the second stage. Subsequently, we run the simulation for about 1.25 solar hours to study the small scale reconnection events during this stage.\\

Periodic boundary conditions are applied in both x and y directions. The upper boundary condition is semi-open, allowing only outflows and prohibiting inflows, with a hot plane set at a temperature of {\bf{$10^{6}$}} K. The bottom boundary permits both outflows and inflows, utilizing an isentropic lower boundary condition. The magnetic fields at the upper boundary are determined by potential field extrapolation. To quickly form a layer with coronal temperatures in the top region, we enabled the hot-plane option. The magnetic fields at the bottom boundary adopt symmetric magnetic components (Open-boundary Symmetric-field, OSb, as described in \cite{Rempel2014}). The state equation used in the simulation is the Uppsala equation \citep{Gudiksen2011}. During our focus period, raw data were saved at 5-second intervals for subsequent analysis of reconnection events.\\

%
%


\section{Results}

%
\subsection{The turbulent reconnection mediated with plasmoids}

After extending the simulation domain and running the MURaM code for over one solar hour, the turbulent convective cells in the convection zone and the irregular movement of granules in the photosphere cause magnetic flux to emerge into the atmosphere, forming complex and diverse magnetic field structures. In some regions, magnetic fields with opposite polarities approach each other and reconnection appears. Due to the high spatial resolution in this simulation, the region between 0 and 4 Mm above the photosphere is filled with many small-scale magnetic reconnection events. Turbulent reconnection mediated with plasmoid instability appears in most of these reconnection events. Small twisted magnetic flux ropes or plasmoids are formed in such events, with sizes typically ranging from a few hundred kilometers to about 1 Mm and occurring as low as around z = 1 Mm. They are ejected downward and upward with the bidirectional flows. Therefore, a portion of the twisted flux ropes have been transported to the higher atmosphere.\\

Figure~\ref{fig:Setup} is a 3D overview of the simulation area at t=4414 seconds. The grayscale image at the bottom represents the longitudinal magnetic field at the solar surface, the colorful lines represent the magnetic field lines in the simulation area, while the red-to-yellow isosurfaces represent the regions where the temperature is higher than 20,000 K. One can find that the local regions have been heated to such a high temperature in many of the small scale reconnection events. Here, we focus on a reconnection event near the left surface as shown in the purple box in Fig.~\ref{fig:Setup}, which is located around x=2 Mm, y=4Mm and extended from z=1Mm to z=5.5 Mm in the z direction (This is our primary region of interest, hereafter referred to "ROI"). This reconnection region shows as a curved, elongated structure resembling a sausage, with an uneven distribution of high-temperature components. The blob-like features are visible, which indicates the occurrences of plasmoid instability in this reconnection event. \\

Figure~\ref{fig:Slices1} displays the distributions of temperature and longitudinal magnetic field (Bz) in the x-z slices at y=4.3 Mm at three different times. The Bz distribution shown in panels b, d, and f indicates that the the local reconnection events are abundant in this plane. Among all of them, the elongated reconnection region between x=0.5 Mm and x=2.5Mm (corresponding to ROI) is notable as shown in panels a,c,e. At t=4301 s, no significant plasmoids appear, and the temperature in the reconnection area does not exceed 20,000 K. At t=4414 s, Fig.~\ref{fig:Slices1}c shows clear plasmoid structures and highly non-uniform temperature distributions in the ROI, with the highest temperature in this region exceeding 90,000 K. At t=4489 s, we can see that the plasmoid structures become not obvious again, and some of the high temperature regions disappear. \\

Figure~\ref{fig:Slices2} further illustrates the presence of plasmoid instability in the ROI with slices cut through different directions and locations. Figure~\ref{fig:Slices2}a displays the distribution of the longitudinal magnetic field at the solar surface, which is primarily influenced by the turbulent movement of granular structures; the magnetic fields converge in the intergranular lanes, with the strongest field strengths exceeding 1500 G. We observe that positive and negative magnetic fields approach each other at many different locations. The ROI is located above the position near x=2Mm, y=4Mm, where exists small scale positive and negative magnetic fields. The distribution of plasma temperature, density, and current density in two x-direction slices (S1, S2) and a y-direction slice (S3) at t=4414 s are presented in Fig.~\ref{fig:Slices2}b-i, these slices passing through the ROI.  The elongated current sheet in ROI is very turbulent and mediated with plasmoids in these planes, the plasmas with different temperatures and densities alternate to appear in the reconnection region. We can see that the cool plasma below 10,000 K can even appear above the hot plasma exceeding 20,000 K, such as the big relatively cooler blob passed by the blue arrow in Figs.~\ref{fig:Slices2}b\&\ref{fig:Slices2}d.  These phenomena are very similar to previous two-dimensional simulations \citep[e.g.,][]{Ni2021,Cheng2024} where plasmoids appeared as magnetic islands, but in the three-dimensional simulation, they correspond to magnetic flux ropes. We can also see that the current sheet extends to higher heights than in previous two-dimensional \citep{Ni2021,Cheng2024} simulations (reaching up to y=6 Mm), with the background plasma density much lower. However, the current sheet is still filled with high-density plasma up to 10$^{19}\cdot$m$^{-3}$, and a large amount of high-density plasma is pushed into the reconnection region or even reach above the current sheet during the magnetic flux emergence process, suggesting that the possible occurrences of EBs at higher heights.\\

The 3D distributions of magnetic field lines and plasmas with higher and lower temperature mixed in the ROI are presented in Fig.~\ref{fig:plasmoid}. Panels a and b show that multiple twisted magnetic flux ropes corresponding to plasmoids are generated in the reconnection region, and those upward moving flux ropes  are unraveling and interacting with the open field. The reddish isosurfaces in panels c and d represent the high temperature regions (above 20,000 K), and the greenish isosurfaces correspond to the low temperature regions (below 8,000 K), we can find that the high and low temperature regions are intertwined within the current sheet. The high temperatures above 20,000 K also appears in reconnection outflow regions. The fragmented high temperature regions in the current sheet and outflows further indicate the appearances of turbulent reconnection. Additionally, supplementary movie shows some twisted magnetic ropes entering higher atmospheric layers with the outflows. Moreover, we find the arch-like magnetic field structures below the reconnection current sheet, which resemble the post reconnection loops, but the temperature in this region does not exceed 20,000 K.\\

\subsection{The Radiation characteristics of reconnection events}

We utilized the RH1.5 code \citep{Pereira2015} to synthesize the spectral line profiles and images in the H$\alpha$ and SiIV pass bands, in order to further study the reconnection activities and more effectively compare them with observations. RH1.5D solves the radiative transfer processes column by column in the three-dimensional simulation region, treating the three-dimensional atmosphere as a 1.5-dimensional distribution to save computational resources. Since computing the Si IV emission requires a complex model of the Si atom, using the RH1.5D code to calculate the radiative characteristics of Si IV 139.4 nm for the entire simulation region involves a substantial computational cost. Therefore, we primarily employ the optically thin assumption combined with the CHIANTI atomic database \citep{Dere1997}  to synthesize the images in Si IV 139.4 nm and EUI (Solar Orbiter/Extreme Ultraviolet Imager)17.4 nm. But the Si IV spectral line profiles are synthesized by using the RH1.5 code. \\

Figsures~\ref{fig:Slices1}, \ref{fig:Slices2} and \ref{fig:plasmoid}  reveal the multi-thermal structures of magnetic reconnection in ROI, the temperature increase in the low-temperature region is only about hundreds to thousands K and their plasma density ranges from 10$^{-10}$ to 10$^{-9}$ g$\cdot$cm$^{-3}$, which are highly conducive for the formation of EBs. The highest temperatures in the current sheet can approach 90,000 K, and the high temperature regions exceeding 20,000 K are very likely to have strong UV emissions. The conditions for the formation of EBs and UV bursts are simultaneously satisfied within a multi-thermal reconnection current sheet. To verify whether both EBs and UV bursts coexist in the same current sheet, we firstly use the RH1.5D code to synthesize the images in the H$\alpha$ wing along z-direction at t=4414 s. Additionally, we also synthesize images in the Si IV 139.4 nm and EUI 17.4 nm by using the optically thin method, with the line-of-sight direction perpendicular to the solar surface (z-direction). The results are presented in Fig.~\ref{fig:syn1}. Panel~a displays a synthetic image near H$\alpha$ -0.5 Å, clearly showing there are many noticeable brightenings in the simulation domain. The UV emission in Si IV 139.4 nm are also significant in many tiny regions as shown in Panel b. However, Panel c indicates that the extreme ultraviolet emissions are not obvious in the whole domain, especially within the ROI. The ROI is within the red dashed box near the left boundary in Panels a and b. \\

Figure~\ref{fig:synbox} shows the zoomed in pictures of the synthetic images in H$\alpha$ wing and Si IV 139.4 nm within the red dashed box plotted in Fig.~\ref{fig:syn1}.  The regions with obvious H$\alpha$ wing emissions are highlighted by the blue isolines in Panel a. The green isolines in Panels a and b are used to mark the regions with significant Si IV emissions. The long curved reconnection region with multiple plasmoids and multi-thermal structures as shown in Figs. \ref{fig:Slices1}, \ref{fig:Slices2} and \ref{fig:plasmoid} is located  from x=0 Mm to x=2.5 Mm and from y=3.5 Mm to y=5 Mm. In this region, one can find that some areas within the blue isolines overlap with or close to the areas within the green isolines, which indicate the possible appearances of UV bursts connecting with EBs. \\

Assuming the red dashed line in Fig.\ref{fig:synbox}a represents the slit of a spectrograph, we synthesized the H$\alpha$ spectral image for the slit as shown in Fig.\ref{fig:spectrohelio}a. Its main features are similar to those observed by \cite{Ellerman}, with no obvious radiations at H$\alpha$ line center but significant emission enhancements in the H$\alpha$ blue and red wings. In Fig.~\ref{fig:spectrohelio}b, we show the temperature distribution in the y-z plane along the red dashed line in Fig.\ref{fig:synbox}, and the gray line in this plane represents the location where the optical depth of H$\alpha$ wing (-0.5 Å) equals unit. Fig.~\ref{fig:spectrohelio}c shows the vertical velocity distribution in the same plane, with the maximum velocity in the \( z \)-direction reaching up to 70~km\,s\(^{-1}\). Fig.~\ref{fig:spectrohelio}d presents the Si~IV emission intensity synthesized using the optically thin approximation. One can see that the regions with strong Si IV emissions in Fig.\ref{fig:spectrohelio}d correspond well to the regions with high temperature plasmas in Fig.\ref{fig:spectrohelio}b. These high temperature plasmas are mostly located below the gray line in Fig.\ref{fig:spectrohelio}b, which indicates they are at a lower chromosphere. \\

We also used the RH1.5D to calculate the averaged Si IV spectral line profile in the small blue box of Fig.\ref{fig:synbox}b, with the line of sight along the z-direction. The result is shown in Fig.\ref{fig:spectrohelio}e. The Si IV 139.4 nm line profile in this area exhibits a distinct double-peaked emission structure, and the spectral line width exceeds 70 km$\cdot$s$^{-1}$. These results match the main observational features of UV bursts. The simulation results also suggest that the reconnection downward outflows can reach a speed of 100 km$\cdot$s$^{-1}$, and upward outflows approximately reach 80 km$\cdot$s$^{-1}$, which is consistent with the recent observational results of small-scale events in the lower solar atmosphere \citep[e.g.,][]{Leenaarts2025}.



The above analyses about the radiation characteristics of the target reconnection event in the ROI further demonstrate that the UV busts and EBs coexist in a single reconnection event, the multi-thermal plasmas with different temperatures caused by turbulent reconnection process contribute emissions in different pass bands.  In addition to this target event, there are many other reconnection events in  the whole simulation domain. Some of them exhibit even stronger radiative signatures. However, this particular event shows more well-identified plasmoid structures.\\

\subsection{The formation height of EBs and UV bursts}

The current sheet presented in Fig.~\ref{fig:Slices2} displays a complex multi-thermal structure. We can see a distinct low temperature (< 10,000 K) plasma blob is located on the right hand side of the curved current sheet,  with its center near x=1.8 Mm and y=3 Mm. The evident high temperature (> 20,000 K) current sheet fragments are located below this cooler plasma blob. However, we also notice that the temperature of this blob is still higher than the surrounding environment. As shown in Fig.~\ref{fig:Halpaline}, we calculated the averaged H$\alpha$ line profiles along the z-direction through this plasma blob (marked by the blue arrows in Fig.~\ref{fig:Slices2}b and d), with specific calculations performed inside the small blue box in Fig.~\ref{fig:Slices2}a. In Fig.~\ref{fig:Halpaline}, the H$\alpha$ line profile shows distinct bulges on both wings deviating from a Gaussian profile, indicating enhanced radiation in the wings. Figure~\ref{fig:Halpaline}b displays the H$\alpha$ line profile after subtracting the nearby background profile, which also shows enhancements in the wings, though the enhancement at the line center is very weak. Therefore, this plasma blob exhibits clear EB characteristics, which is located above some high-temperature areas of the current sheet. These results imply that in a magnetic reconnection event where both EBs and UV bursts occur simultaneously, the much cooler EBs can locate at a higher height than the hot UV emission.\\

Such kind of reconnection events with multi-thermal structures frequently appear in the whole simulation, and the hot plasmas with strong UV emissions can locate at a very low height. Fig.\ref{fig:slices3} shows a slice of another magnetic reconnection event at $t=4548$ s, with Panels a, b, c, and d displaying the longitudinal magnetic field (Bz), temperature distribution, longitudinal velocity (Vz), and density distribution, respectively. The figures reveal that high-temperature plasma exceeding 20,000 K appears in a typical low chromospheric environment at about 0.7 Mm above the solar surface. The density in most of the reconnection region is significantly higher than the background, except for the high-temperature area. Bz at the bottom of this reconnection region is about 300 G, with a total magnetic field strength of about 500 G. Panel e shows the Si IV spectral line profile along the z-direction calculated by RH1.5D for this region, which also exhibits a double-peaked structure. The Doppler width of the line is around 50 km$\cdot$s$^{-1}$, and its radiation intensity is significantly higher than that shown in Fig.\ref{fig:spectrohelio}e, which is attributed to the considerably higher density in this reconnection event. Panel f shows the results of the synthetic image in Si IV, viewed along the y-direction, passing through the high-temperature plasma area of the reconnection. The reconnection current sheet shows a distinct response in the Si IV band, presenting as a slender bright band structure, similar to the flame-like structures commonly observed. Such a result demonstrates that UV bursts can appear in the lower chromosphere, as long as the reconnection magnetic fields are strong enough, which is consistent with the previous 2D simulation results \citep{Ni2016,Ni2018,Ni2022,Cheng2024}.\\

\section{Conclusions and Discussions}

We conducted a high-resolution, three-dimensional RMHD simulation using the MURaM code, covering the upper convection zone to the lower corona. In this simulation, the magnetic field is shaped by the irregular motions of convective cells and photospheric granules, and these magnetic fields rise from the convection zone into the solar atmosphere, subsequently forming various dynamic structures. High-density plasma (>10$^{19}$ m$^{-3}$) in the lower atmosphere also ascends to higher layers with the emerged magnetic field. Throughout the simulation, frequent appearances of magnetic structures with opposite polarities were observed, leading to many small-scale magnetic reconnection events. The high resolution enabled the capture of fine plasma structures within reconnection current sheets and facilitated observations of nonuniform distributions of density and temperature. We tracked the evolution of a target region (ROI) and used the RH1.5D code and the optically thin radiation approximation model to synthesize radiation images and spectral line profiles for H$\alpha$, Si IV 139.4 nm, and EUI 17.4 nm. These led to several important conclusions:\\

\begin{enumerate}

	\item In the simulation, most small-scale reconnection lead to the formation of twisted magnetic flux ropes in the reconnection regions, resembling plasmoids. Some of these flux ropes, driven by bi-directional flows, move upward away from the solar surface, carrying twisted magnetic fields into the upper layers of the solar corona. Additionally, the magnetic topology of the reconnection process seen in the targeted event resembles those found in large-scale solar flare models.\\
	
	\item The turbulent magnetic reconnection process leads to highly nonuniform distributions of density and temperature in the reconnection regions, where plasmas with different temperatures alternately appear in space. A part of plasma in these reconnections is heated to above 20,000 K, with peak temperatures reaching 90,000 K. This high-temperature plasma generates strong Si IV ultraviolet radiation. In contrast, cooler regions of the reconnection processes cause a significant enhancement in the H$\alpha$ line wing radiation. The synthesized spectral line profiles and imaging results in the H$\alpha$ and Si IV bands through the targeted reconnection region indicate the coexisting of EBs and UV bursts during this magnetic reconnection process.\\
	
	\item In the turbulent magnetic reconnection events involving a mixture of high and low temperatures, cool plasma displaying characteristics of EBs can be found above hot plasma, reaching altitudes more than 2Mm above the solar surface. Meanwhile, some high-temperature plasma featuring characteristics of UV bursts can extend downwards to the lower chromosphere, approximately 0.7Mm above the solar surface.\\
	
\end{enumerate}

Our analysis of the targeted event suggests that the multi-thermal turbulent current sheet structures caused by plasmoid instability can plausibly explain the formation mechanisms of UV bursts connecting with EBs, further validating the model we previously proposed in two-dimensional simulations \citep{Ni2021,Cheng2024}. The reconnection structures presented in the three-dimensional simulation are significantly more complex. We should also point out that the plasmoid instability can only occur when the aspect ratio of the current sheet or the Lundquist number exceeds a critical value \citep[e.g.,][]{Ni2012, Leake2012}. The lower solar atmosphere is normally very dynamic and the current sheet aspect ratio can vary with time, then the plasmoid instability can possibly only last for a while and then disappear.

The magnetic field strength in the target event region approximately  ranges from 150 G to 250 G, and it decreases with height, which is similar to our previous 2D work in \cite{Cheng2024}. Such a curved current sheet extended from the lower atmosphere to the higher altitude is significantly different from the horizontal current sheet in \cite{Ni2022}. The initial uniform plasma density in \cite{Ni2022} is about 10$^{21}$ m$^{-3}$ (comparable to the density in the solar temperature minimum region), and it is more than 10 times higher than that in the target reconnection region in this work. Therefore, the higher strength of reconnection magnetic field (>$\sim$ 500 G) is required to heat the denser plasma there to a temperature above 20,000 K in \cite{Ni2022}.


The simulation results also indicate that the turbulent reconnection mediated by plasmoids is widespread in the lower solar atmosphere and may play an important role in non-homogeneously heating the atmosphere. In particular, if elongated current sheets with lengths of 2–3 Mm or more can persist for some time under low plasma-beta conditions, as is readily achieved in this simulation. Recent high-resolution observations have revealed the complex nonuniform thermal structure of the lower solar atmosphere \citep[e.g.,][]{Peter}, where the turbulent reconnection mediated by plasmoids may play a significant role. As we know, the plasmoid-like structures have been frequently observed in the large-scale current sheet related to solar flares. In our target reconnection event, the magnetic topology resembles that of the standard flare model, suggesting potential similarities between these events at different scales. Despite differences in formation locations and scales, they may share similar triggering mechanisms and an underlying physical model.

Finally, we point out some of the limitations in our study. For instance, the height in the z-direction is set to 8 Mm above the solar surface. As the magnetic flux emergence process continues, a substantial amount of high-density plasma is transported to higher altitudes, pushing the original coronal material out through the upper boundary. This results in only a small amount of coronal material near the upper boundary of the simulation area. In future work, it will be necessary to optimize this aspect to ensure the simulation region encompasses a more extensive coronal plasma environment, better modeling the actual conditions of the Sun.\\

\begin{acknowledgements}
This research is supported by the Strategic Priority Research Program of the Chinese Academy of Sciences with Grant No. XDB0560000; the National Key R$\&$D Program of China No. 2022YFF0503003 (2022YFF0503000); the National Key R$\&$D Program of China No. 2022YFF0503804(2022YFF0503800); the NSFC Grants 12373060 and 11933009; the outstanding member of the Youth Innovation Promotion Association CAS (No. Y2021024 ); the Basic Research of Yunnan Province in China with Grant 202401AS070044; the Yunling Talent Project for the Youth; the Yunling Scholar Project of the Yunnan Province and the Yunnan Province Scientist Workshop of Solar Physics; the International Space Science Institute (ISSI) in Bern, through ISSI International Team project \# 23-586 (Novel Insights Into Bursts, Bombs, and Brightenings in the Solar Atmosphere from Solar Orbiter); Yunnan Key Laboratory of Solar Physics and Space Science under the number 202205AG070009; The numerical calculations and data analysis have been done on Max Planck Computing and Data Facility and on the Computational Solar Physics Laboratory of Yunnan Observatories.

\end{acknowledgements}

%
%

\end{document}